\documentclass[twocolumn,showpacs,preprintnumbers,amsmath,amssymb,prb]{revtex4}

 \usepackage[dvips]{graphicx}
 \usepackage[dvips]{graphics}
 \usepackage{color}

\begin{document}

\title{Characterization of the Intra-Unit-Cell magnetic order in  $\rm Bi_2Sr_2CaCu_2O_{8+\delta}$}

\author{L. Mangin-Thro}
\affiliation{Laboratoire L\'eon Brillouin, CEA-CNRS, CEA-Saclay, 91191 Gif sur Yvette, France}

\author{S. De Almeida-Didry}
\affiliation{Laboratoire L\'eon Brillouin, CEA-CNRS, CEA-Saclay, 91191 Gif sur Yvette, France}
\affiliation{Universit\'e Fran\c cois Rabelais de Tours, CNRS, CEA, ENIVL, GREMAN UMR 7347, IUT de Blois  15 rue de la chocolaterie 41000 Blois, France}

\author{Y. Sidis}
\email{yvan.sidis@cea.fr}
\affiliation{Laboratoire L\'eon Brillouin, CEA-CNRS, CEA-Saclay, 91191 Gif sur Yvette, France}

\author{F. Giovannelli}
\affiliation{Universit\'e Fran\c cois Rabelais de Tours, CNRS, CEA, ENIVL, GREMAN UMR 7347, IUT de Blois  15 rue de la chocolaterie 41000 Blois, France}


\author{I. Laffez-Monot}
\affiliation{Universit\'e Fran\c cois Rabelais de Tours, CNRS, CEA, ENIVL, GREMAN UMR 7347, IUT de Blois  15 rue de la chocolaterie 41000 Blois, France}

\author{P. Bourges}
\email{philippe.bourges@cea.fr} \affiliation{Laboratoire L\'eon Brillouin, CEA-CNRS, CEA-Saclay, 91191 Gif sur Yvette, France}

\date{\today}

\pacs{PACS numbers: 74.25.Ha, 74.25.Dw, 74.72.Kf, 78.70.Nx}

\begin{abstract}
As in $\rm YBa_2Cu_3O_{6+x}$ and $\rm HgBa_2CuO_{8+\delta}$, the pseudo-gap state in $\rm Bi_2Sr_2CaCu_2O_{8+\delta}$ is characterized by the existence of an intra-unit-cell magnetic order revealed by polarized neutron scattering technique. We  report here a supplementary set of polarized neutron scattering measurements for which the direction of the magnetic moment is determined and the magnetic intensity is calibrated in absolute units. The new data allow a close comparison between bilayer systems $\rm YBa_2Cu_3O_{6+x}$  and $\rm Bi_2Sr_2CaCu_2O_{8+\delta}$ and rise important questions concerning the range of the magnetic correlations and the role of disorder around optimal doping.

\end{abstract}

\maketitle


\section{\label{Intro} Introduction}

Recent ultrasound measurements in $\rm YBa_2Cu_3O_{6+x}$ (Y123) have  provided conclusive evidence that this system undergoes a true phase transition at a temperature $\rm T^{\star}$ upon entering the pseudo-gap (PG) state \cite{Shekhter-nature}. This study gives a thermodynamic signature of a PG phase as a new state of matter distinct from the superconducting (SC) state. The elastic response exhibits an anomaly associated with the electronic transition \cite{Shekhter-nature} which takes place at the same temperature where an intra unit cell (IUC) magnetic order develops according to polarized neutron scattering experiments performed in Y123 system \cite{Fauque,Sidis,Mook,CC-review,Baledent-YBCO,CC-review2}. This magnetic order breaks time reversal symmetry but preserves lattice translation invariance. At variance with a  simple ferromagnetic order, the unit cell does not exhibit any net magnetization. Such an order can be found as soon as staggered magnetic moments develop within the unit cell (in that case, one usually speaks about a q=0 antiferromagnetic order). Detailed macroscopic magnetic susceptibility measurements in Y123\cite{Leridon} indicate anomalies at the same temperature as the IUC magnetic order, confirming the magnetic nature of the transition. In addition at a slightly lower temperature, high resolution magneto-optic measurements\cite{xia08} show an anomalous Kerr effect in the PG state in the same Y123 system. First observed in Y123, the IUC magnetic order has been reported in three other cuprates families: $\rm HgBa_2CuO_{4+\delta}$ (Hg1201) \cite{Li-Nature,Li-PRB}, $\rm La_{2-x}Sr_xCuO_4$ (La214) \cite{Baledent-LSCO} and $\rm Bi_2Sr_2CaCu_2O_{8+\delta}$ (Bi2212) \cite{DeAlmeida-PRB12}. However,  the observation of such a magnetic order by polarized neutron technique has not been corroborated by magnetic local probe measurements: no indication of a magnetic transition has been observed so far using either nuclear magnetic resonance (NMR) techniques \cite{strassle,mounce} or zero-field muon-spin rotation ($\mu$SR) technique \cite{sonier,uemura}.
To reconcile, bulk polarized neutron scattering and local probe measurements, one can speculate \cite{Fauque,RFIM} that the IUC magnetic order develops within  finite size domains and may fluctuate at a  characteristic time-scale intermediate between the neutron time-scale  (10$^{-11}$ sec) and the local probes time-scales (typically 10$^{-6}$-10$^{-8}$ sec) explaining why these magnetic correlations cannot be detected in $\mu$SR\cite{sonier,uemura} and NMR experiments \cite{strassle,mounce} as they remain dynamic at their time-scales.

The IUC magnetism could be induced by counter-circulating loop currents within the unit cell of $\rm CuO_2$ planes, as predicted by C. M. Varma in his Loop-Current (LC) theory of the PG state \cite{RFIM,Varma06}. Supporting this mean-field theory, the study of the phase diagram of an effective three orbital model of cuprates using Variational Monte-Carlo calculations has recently shown the stability of the LC state in the thermodynamic limit \cite{Weber13}. But this theory faces a serious problem, since a q=0 electronic instability is not expected, in principle, to open a gap in the charge excitation spectrum. Still within the LC model, a possible solution around this conundrum has been recently proposed \cite{RFIM} as the ground state exhibits four degenerate discrete LC configurations. In the presence of disorder, fluctuating finite-size domains occur between these equivalent configurations. Within a random field approximation, the disorder gives rise to a "central peak", directly observable in neutron diffraction as a Bragg peak. However, these magnetic fluctuations would not be observable by local probes due to motional narrowing \cite{RFIM}. Meanwhile, a singular forward scattering of fermions for large correlation lengths induces a pseudo-gap in the single-particle spectral function near the chemical potential \cite{RFIM}.

Alternative models have been recently developed to account for the neutron signal. First, still within the loop current approach, interesting proposals have been built where the currents flow around the full CuO$_6$ octahedra \cite{Weber09,Lederer} but still having the same symmetry as the original LC model \cite{Varma06}. Counter-intuitively, this add-on actually does not improve significantly the description of the neutron intensity. Interestingly, chiral order parameter based on the LC model has been also developed to explain the unusual polar Kerr effect\cite{yakovenko}. Strictly speaking, this proposal of chiral order does not also correspond to the experimental neutron data as it would produce extra Bragg peaks along c*, the direction perpendicular to the CuO$_2$ plaquette. For instance, in Y123, it would produce a doubling of the unit cell along c* meaning an extra Bragg peak at L=1/2 which is clearly not observed in Y123  \cite{Mook}. On a different perspective, Ising-like orbital magnetic moments of oxygen atoms have been proposed to occur as a possible ground state of doped CuO$_2$ plaquette \cite{Moskvin}. Such intra-CuO$_2$ plaquette staggered order exhibits the same symmetry as the spin moments at the oxygen sites\cite{Fauque} and can as well account for the neutron intensity. Recently, a new model for the PG state has been proposed based on a composite charge stripe order \cite{Chubukov}. This model proposes the PG physics is controlled by a stripe charge order parameter with two components: one is an incommensurate density variation, another is an incommensurate current. Such an order breaks time reversal symmetry and generates loop currents. Interestingly, the LC order can set up at a higher temperature than the charge density wave, when the density and the current components form a composite order with zero total momentum. This model suggests that it might exist different ways to account for the existence of IUC order in the phase diagram of cuprates.

To check the relevance of these scenarios, one needs to study the interplay between the opening of a pseudo-gap in the charge excitation spectrum, the observation of an IUC order and the disorder in cuprate materials. To this respect, the Bi2212 cuprate family is of particular interest. As the Y123 system, Bi2212 is a bilayer cuprate, but unlike Y123, its high surface quality allows a detailed study of the PG in the charge excitation spectrum on a size of a few micrometers using angle resolved photoemission (ARPES) \cite{Vishik} and electronic Raman spectroscopy \cite{Sacuto} (ERS) or at the atomic scale using scanning tunneling microscope (STM) spectroscopy \cite{Lawler,Hoffman}. In Bi2212,  two distinct techniques, polarized neutron scattering \cite{DeAlmeida-PRB12}, on the one hand, and  circularly polarized angle resolved photoemission (ARPES) \cite{Kaminski-Dichroism,Varma-Simon}, on the other hand, provide evidence for a time reversal breaking state below the PG temperature $\rm T^{\star}$. The temperature dependencies of the magnetic neutron scattering intensity and dichroic effect in ARPES display a striking similarity \cite{CC-review2}. The existence of an IUC order in the PG phase of Bi2212 system is also supported by the observation of an anisotropic electronic density of state reported by STM spectroscopy \cite{Lawler}. Furthermore, the existence of nanoscale electronic inhomogeneities can be imaged using STM spectroscopy \cite{Hoffman}. Interestingly, specific signatures of orbital loop currents have been predicted\cite{Nielsen} in the spatially resolved local density of states of STM.

\begin{figure}[t]
 \includegraphics[width=7cm,angle=0]{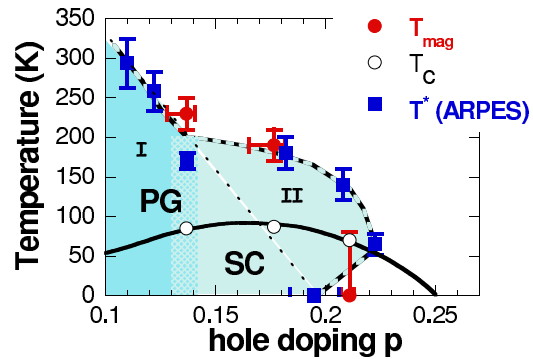}
\caption {
(color online) Hole doping (p) dependencies of the pseudo-gap (PG) temperature $\rm T^{\star}$ measured by ARPES in Bi2212 system \cite{Vishik} and the characteristic temperature associated with the appearance of the IUC magnetic phase $\rm T_{mag}$, as determined by polarized neutron scattering technique. In the PG state, following the ARPES measurements \cite{Vishik}, one can distinguish two regimes, labelled I and II, as explained in the text.
}
\label{PRB2-Fig0}
\end{figure}

In Y123 \cite{Fauque} and Hg1201 \cite{Li-Nature},  the variation as a function of the hole doping of the magnetic ordering temperature $\rm T_{mag}$ matches  the evolution of $\rm T^{\star}$ determined from resistivity measurements. Both temperatures decrease linearly upon increasing the hole doping. They tend to vanish around a critical hole doping $\rm p_c \sim 0.19$, considered as the end point of the PG state according to thermodynamic measurements. In Bi2212,  the hole doping dependence of $\rm T_{mag}$ \cite{DeAlmeida-PRB12} fits nicely the one of $\rm T^{\star}$ deduced from ARPES measurements \cite{Vishik}, as shown in Fig.~\ref{PRB2-Fig0}. Actually, two different regimes have been inferred from these ARPES measurements \cite{Vishik}: $\rm T^{\star}$ first decreases linearly with the doping (regime I), but upon approaching the optimal doping (p=0.16) it exhibits a plateau and then steeply vanishes at larger doping (regime II) with a reentrant behaviour. In Y123 and Hg1201, $\rm T_{mag} \simeq \rm T^{\star}$ has been found to decrease linearly at low doping corresponding to the regime I. The crossover between regimes I and II is likely to take place around $\rm p\sim 0.13-0.14$ (Fig.~\ref{PRB2-Fig0}).

In this article, we present a polarized neutron scattering study of the underdoped Bi2212 sample, located at the crossover between regimes I and II in Fig.~\ref{PRB2-Fig0}. The existence of the magnetic IUC order for that sample has been established in a previous study \cite{DeAlmeida-PRB12}. The purpose of the present study is to provide a more quantitative description of this order. We report a full polarization analysis which gives access to the orientation of the magnetic moments. Indeed, in the LC theory, orbital magnetic moment should be perpendicular to the $\rm CuO_2$ planes, whereas the observed magnetic moments display a non negligible planar component in Y123, Hg1201 and La214 \cite{Fauque,Mook,CC-review,CC-review2}. The determination of the magnetic moment orientation in Bi2212 has not been addressed so far. To complete the description of the IUC magnetic order, we extend our polarized neutron scattering study to an overdoped Bi2212 sample  in regime II (see Fig.~\ref{PRB2-Fig0}). The calibration of the observed magnetic neutron scattering intensity in absolute units allows a quantitative comparison between Y123 and Bi2212 bilayer systems. This comparison suggests that the IUC magnetic order still shows up at rather high temperature upon increasing the hole doping, but the IUC magnetic correlation length is likely to shorten when passing from regime I to regime II.

\begin{figure}[t]
 \includegraphics[width=5.5cm,angle=0]{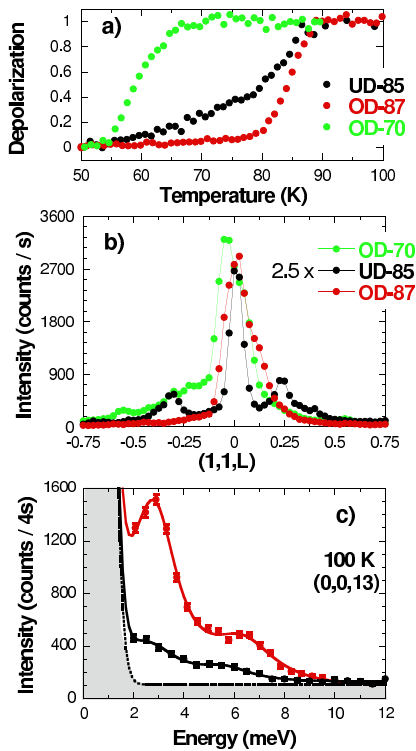}
\caption {
(color online) a) Characterization of the $\rm T_c$ onset measured by neutron depolarization technique for 3 samples: UD-85 (black)\cite{DeAlmeida-sample,DeAlmeida-PRB12}, OD-87 (red) \cite{Fauque-Bi2212}, OD-70 (green) \cite{Capogna-Bi2212}. While the superconducting transition is rather steep for the overdoped samples,  it becomes  much smoother for the underdoped sample, a characteristic feature of the underdoping.
b) Rocking scans across the Bragg peak $\rm {\bf Q}$=(1,1,0). The measured intensity for sample UD-85 is multipled by a factor of 2.5, for sake of comparison with the other samples. c) Low energy phonon measurements at the wave vector $\rm {\bf Q}$=(0,0,13) on sample UD-85 and OD-87. The shaded area corresponds to the incoherent elastic scattering on top of the nuclear background.
}
\label{PRB2-Fig1}
\end{figure}

\section{\label{Exp} Experimental details}
\subsubsection*{\label{sample} Sample preparation and characterization }

Measurements are performed on underdoped (UD) Bi2212 single crystals. The synthesis is carried using the travelling solvent floating zone technique (TSFZ) in air \cite{DeAlmeida-sample}. Three large single crystals are first extracted from the as-grown rod. The composition homogeneity and the bulk crystal quality are provided by EDX and neutron scattering diffraction studies. The as-grown crystals are actually weakly overdoped (OD) with a nominal superconducting critical temperature $\rm T_c$ of 87 K. The underdoping of the samples is then achieved using a post-annealing treatment of 100 h under reduced oxygen atmosphere $\rm P(O_2)$=0.1 atm at 450$\rm ^{o}$C, yielding an average onset $\rm T_c$ of 85 K \cite{DeAlmeida-sample}. $\rm T_c$ is determined by magnetization measurements and neutron depolarization technique (Fig.~\ref{PRB2-Fig1}.a). In addition to the $\rm T_c$ reduction, the loss of mass of the sample, on the one hand, and the value of the $\rm \bf{c}$ axis parameter, on the other hand, match those reported in the literature for UD Bi2212 single crystals obtained using a similar method \cite{Liang-sample}. In order to further establish the underdoped nature of our samples,  smaller samples from the same batch, are further studied in the SC state by ERS in both $\rm B_{1g}$ and $\rm B_{2g}$ channels. According to previous ERS measurements in UD-Bi2212 samples \cite{Blanc-2010}, the hallmark of the SC has to be washed out in the $\rm B_{1g}$ channel, while it remains observable in  the $\rm B_{2g}$ channel: a characteristic property recovered in our samples \cite{Gallais}.

Once the three single crystals are co-aligned, the total sample volume reaches  a nominal value of 330 mm$\rm ^{3}$. However, the actual sample volume that really contributes to the neutron scattering measurement is significantly lower. To estimate the useful sample fraction, our UD-85 sample is compared with two other Bi2212 samples used in previous neutron scattering studies : the samples OD-87 \cite{Fauque-Bi2212} and OD-70 \cite{Capogna-Bi2212}. Fig.~\ref{PRB2-Fig1}.a shows the bulk $\rm T_{c}$ of all samples, measured by neutron depolarization technique. The OD-87 sample is a rod-like sample of 300 mm$\rm ^{3}$ with a main single crystal of 250 mm$\rm ^{3}$ and a smaller one of 50 mm$\rm ^{3}$ shifted at 2$\rm ^{o}$. The OD-70 sample is an array of single crystals glued on 3 Al plates. Their  quality was initial checked using X-ray Laue diffraction measurements. The sample mass was first estimated to be twice larger than that of the OD-87 sample \cite{Capogna-Bi2212}. However, the bulk sample quality was further crosschecked using neutron diffraction measurements and almost half of the sample had to be removed, yielding a final sample volume of  $\sim$300 $\rm mm^{3}$.

The average structure of the Bi2212 compound is usually described within an orthorhombic unit-cell a$\simeq$ b $\simeq$
5.4 \AA\  and c$\simeq$  30.9 \AA. It exhibits a strong one-dimensional incommensurate modulation with the wave vector
${\bf q_s} = 0.21 {\bf b^*} + {\bf c^*}$\cite{janine}. For a sake of comparison with our previous studies of the IUC magnetic order, we rather adopt the tetragonal lattice unit cell with a=b $\simeq$  3.82 \AA, turned by $\rm 45^{o}$ within the ab plane from the orthorhombic lattice. Within all this paper, $ \rm \bf{Q}$ is then given in reduced lattice units $ \rm  ( \frac{2 \pi}{a}, \frac{2 \pi}{b}, \frac{2 \pi}{c}) $, using tetragonal notations $\rm a= b=$ 3.82 ~\AA  ~and c=30.87 ~\AA.

The samples UD-85, OD-87 and OD-70 were aligned in  the [110]/[001] scattering plane  and characterized on the triple axis spectrometer G43 located in the guide hall of the Orph\'ee reactor (Saclay) with a final wave vector $\rm k_f$=1.97 ~\AA$\rm ^{-1}$. Fig.\ref{PRB2-Fig1}.b shows transverse scans around $\rm {\bf Q}$=(1,1,0). The measured  Bragg peak intensity for sample UD-85 is $\sim$ 2.5 times weaker than for the two other samples, whose nuclear Bragg intensities are similar. Likewise their mosaic is about twice larger than that of the UD-85 sample. Considering the integrated Bragg intensity, one can therefore conclude that the actual volume of the UD-85 sample is typically 5 times weaker than the two other samples. Additional low energy phonon measurements at the wave vector $\rm {\bf Q}$=(0,0,13) were performed using the triple axis spectrometer 2T (Orph\'ee reactor-Saclay) on the UD-85 and OD-87 samples (Fig.\ref{PRB2-Fig1}.c). These measurements  also confirm that the UD-85 sample has an actual volume about 5 times smaller than the OD-87 sample. Both elastic and inelastic neutron scattering measurements indicate that the actual volume of  the UD-85 sample  is likely to be only $\sim$50 $\rm mm^{3}$, smaller than the nominal volume due to distribution of grains in the crystal.

\subsubsection*{\label{Polar} Polarized neutron scattering experiment}

For the neutron scattering measurements, the samples are co-aligned using different methods:  the samples are either glued on an Al-plate or wrapped on Al-foils and attached on thin Al rods. Using these methods, we can ensure the reproducibility of the neutron scattering measurements independently from a variation of the magnitude of the background associated with the presence or absence of glue. The samples are attached on the cold head of a 4K-closed cycle refrigerator and aligned in the [100]/[001] scattering plane (tetragonal notations), so that transferred wave vectors $ \rm \bf{Q}$ of the form (H,0,L) are accessible.

Polarized neutron scattering measurements are performed on the cold neutron triple-axis spectrometer 4F1 at reactor Orph\'ee in Saclay (France). The polarized neutron scattering set-up is similar to the one used in previous experiments on the same topic \cite{Fauque,Sidis,Mook,Baledent-YBCO,Li-Nature,Li-PRB,Baledent-LSCO,CC-review}: the incident neutron beam is polarized using a polarizing super-mirror (bender) and the polarization of the scattered beam is analysed using a Heusler analyzer. Standard XYZ-Helmholtz coils guide the neutron spin polarization on the sample. The experimental set-up further includes on the incoming neutron beam a Mezei flipper for flipping the neutron spin direction and a pyrolytic graphite filter in front of the bender to eliminate high order harmonics. For the polarized diffraction measurements, the incident and final neutron wave vectors are set to $k_I$=2.57~\AA$^{-1}$.

\begin{figure}[t]
\includegraphics[width=6.5cm,angle=0]{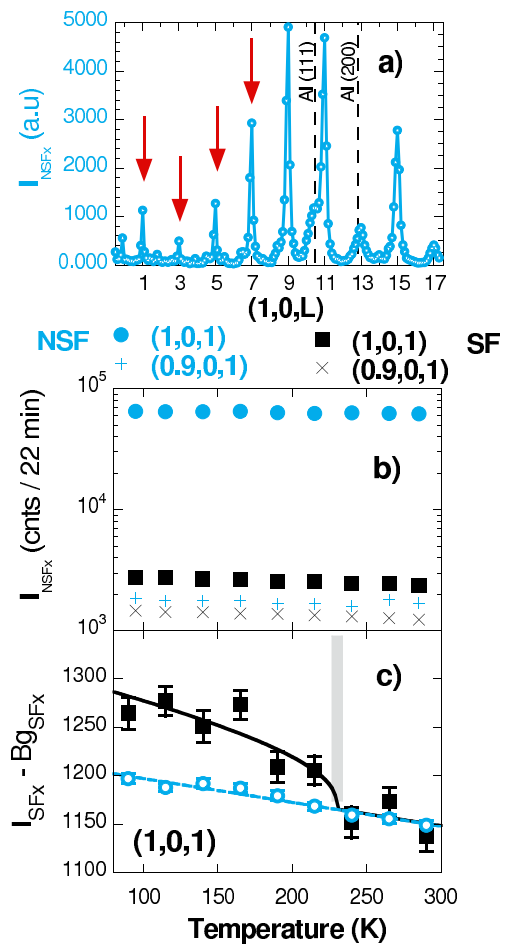}
\caption {
(color online) a) Variation of the nuclear Bragg intensities as a function of {\bf Q}=(1,0,L) in the UD-85 sample. The measurement is performed at room temperature in the NSF channel with the neutron polarization Hx. The nuclear Bragg reflections are located at odd integer L values. Contaminations from Al powder lines also show up at non integer L values and large wave vectors. b) Temperature dependencies of the raw  neutron scattering intensities measured at the Bragg peak {\bf Q}=(1,0,1) (full symbols) and at {\bf Q}=(0.9,0,1) (crosses), the background position, in the NSF channel (light blue) and in the SF channel (black). c) Temperature dependencies of intrinsic Bragg scattering (background subtracted) at {\bf Q}=(1,0,1) in the SF channel (black) and in the NSF channel (light blue) divided by a T-independent bare flipping ratio $\rm FR^o_x$= 52.6 ($\rm 1/FR^o_x$=0.019). The solid lines are discussed in the text.
}
\label{PRB2-Fig2}
\end{figure}

The scattered intensity at a given wave vector $\bf{Q}$ is systematically measured in the spin-flip (SF) and non-spin-flip (NSF) channels, with 3 orthogonal neutron spin polarizations: Hx, Hy, Hz. For Hx and Hy, the neutron spin polarization is respectively parallel and perpendicular to $\bf{Q}$ in the scattering plane. For Hz, the neutron spin polarization is perpendicular to both 
$\bf{Q}$ and the scattering plane. In the rest of the manuscript, the indices SF, NSF and $\alpha$={x,y,z} indicate to which channel and to what kind of polarization the measurements correspond.

In unpolarized neutron diffraction, the measured magnetic intensity is proportional  to $\rm |\bf{M}_{\perp}|^2$, where  $\rm \bf{M}_{\perp}$ stands for the magnetic component of the ordered moment perpendicular to $\bf{Q}$. In polarized neutron diffraction,
the measured magnetic intensity is proportional to  $\rm |\bf{\sigma}.\bf{M}_{\perp}|^2$ where $\bf{\sigma}$ represents the Pauli matrices with $\sigma_z$ defined along the neutron spin polarization $\rm H \alpha$. Therefore, only the $\rm \bf{M}_{\perp}$ component perpendicular to the neutron spin polarization $\rm H \alpha$ contributes to the magnetic intensity in the SF channel, whereas the remaining component contributes to the intensity measured in the NSF channel. As a consequence, the full magnetic intensity always appears in the SF channel for polarization Hx. Likewise, in absence of chirality, the magnetic intensity measured in the SF channel for Hx has to be equal to the sum of the magnetic intensities measured in the SF channels for Hy and Hz, which can be identified as a "polarization sum rule".

\section{\label{results} Experimental results}
\subsubsection*{\label{Order} Evidence for a magnetic order in the underdoped regime }

Following previous studies in bilayer compound Y123 \cite{Fauque,Mook,Baledent-YBCO} and  monolayer compound Hg1201 \cite{Li-Nature,Li-PRB}, the search for a long range magnetic order in the PG phase is performed on the Bragg reflections (1,0,L) with integer L values. It should be noted that, although the in plane (1,0) direction differs from the (0,1) one due to the orthorhombic structure of Bi2212, we did not observe any noticeable difference in our results  in Bi2212 between both directions. We therefore consider here both directions as equivalent.

 For a polarization $\rm H{\alpha}$, the scattered intensity in the SF channel on a Bragg reflection ($\rm I_{SF\alpha}$) is dominated by the leakage of the NSF intensity  into the SF channel, whose magnitude gives the bare flipping ratio ($\rm FR^o_{\alpha}(T)$), characterizing the neutron beam polarization quality and statibility. On top of this signal, the intrinsic magnetic response ($\rm I_{mag\alpha}$), of much weaker intensity in the present case, can develop once a magnetic order settles in below a certain temperature. The scattered intensity in the SF channel then reads:
\begin{equation}
I_{SF\alpha}=  I_{NSF\alpha} / FR^o_{\alpha}(T) + I_{mag \alpha}
\label{I-SF}
\end{equation}
For the polarized neutron scattering measurement in bilayer Bi2212 system, one needs to pay attention to the three following points. First of all, its crystal structure belongs to Bb2b space group: the (1,0,L) Bragg reflections are  therefore observable for odd L values (Fig.~\ref{PRB2-Fig2}.a) only. Next, the neutron beam is quickly depolarized for Bi2212 sample when entering the SC state, prohibiting the identification of any magnetic signal below $\rm T_c$. Finally, in Eq.~\ref{I-SF}, I stands for the intrinsic Bragg intensity, i.e the raw scattered Bragg intensity to which a background (Bg) is removed. In most cases, Bg is sufficiently weak and rather temperature independent to be ignored. This approximation does not hold for Bi2212, as illustrated in Fig.~\ref{PRB2-Fig2}.b. The figure reports the temperature dependencies of the scattered intensities at the Bragg position (1,0,1) and (0.9,0,1) in SF and NSF channels for polarization Hx. The measurements away from the Bragg reflection allow one to estimate the magnitude and the temperature dependence of the background in both SF and NSF channels. The temperature dependencies of $\rm Bg_{SF,x}$ and $\rm Bg_{NSFx}$ are quite similar. $\rm I_{NSFx}$ is  about 2 orders of magnitude larger than $\rm Bg_{NSFx}$, whereas $\rm I_{SFx}$ and $\rm Bg_{SFx}$ are of the same order of magnitude.

In order to properly determine the variation of the Bragg intensity in the SF channel, the background contribution measured at
(0.9,0,L) has to be systematically subtracted from the scattered intensity at (1,0,L). It is worth pointing out that the background does not depend on the neutron spin polarization, indicating that it is free from any magnetic scattering (at least within the experimental accuracy of the present experiment). Once the Bg is subtracted (Fig.~\ref{PRB2-Fig2}.c), one can observe an enhancement of $\rm I_{SFx}$ below $\rm T_{mag}\sim$230 K at $\rm \bf{Q}$=(1,0,1), indicating the appearance of a magnetic order in our underdoped Bi2212 sample. The magnetic signal displays a characteristic T-dependence $\rm \propto (1- \frac{T}{T_{mag}})^{2 \beta}$ with {$\beta=0.2$}. Note that the same power law is used to fit the T-dependence of the magnetic signal hereafter.

\begin{figure}[t]
\includegraphics[width=6.5cm,angle=0]{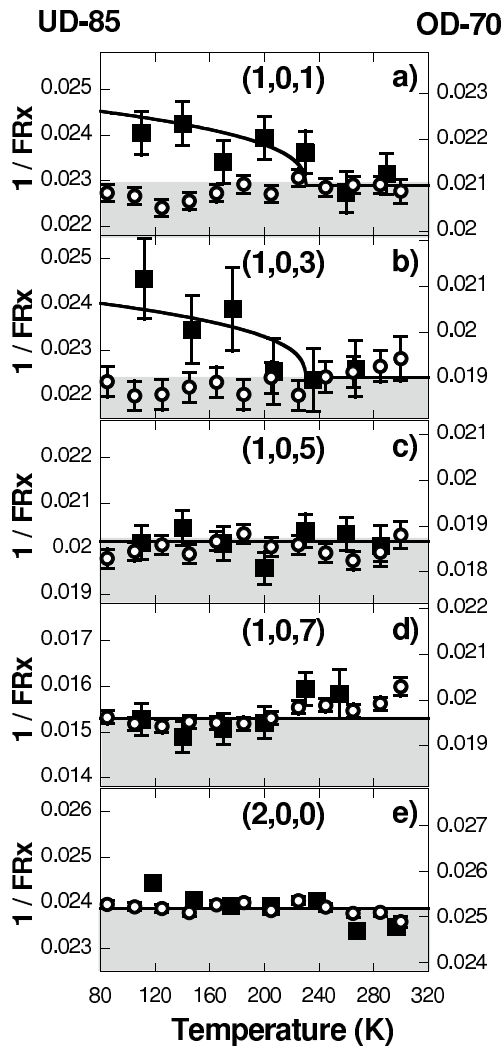}
\caption {
(color online) a-d) Temperature dependencies of $\rm 1/FR_x$ at $\rm {\bf Q}$=(1,0,L) and (2,0,0) for the samples:  UD-85 (left axis - full squares) and OD-70 (right axis - open circles). Solid lines are guides to the eye.
}
\label{PRB2-Fig3}
\end{figure}

\subsubsection*{\label{Momentum} Momentum dependence}

For the comparison of  the same measurements on different samples or on the same sample but obtained during different experiments, it is quite convenient to use the Bragg intensity in SF channel normalised by the Bragg intensity in the NSF channel: this gives the inverse-flipping ratio, $\rm 1/FR(T)$. Using this normalization, Eq.~\ref{I-SF} turns into:
\begin{equation}
1/FR_{\alpha}(T)= 1/FR^o_{\alpha}(T) + I_{mag \alpha} / I_{NSF}
\label{InvFR}
\end{equation}
It is worth noticing that the Bragg intensity measured in the NSF channel is found experimentally independent of the neutron spin polarization, $\rm H{\alpha}$, within the experimental accuracy. The magnitude of $\rm FR^o_{\alpha}$ is given by a measurement at high temperature above the ordering temperature. Its temperature dependence is further determined by an extra measurement at (2,0,0), i.e at large $|\bf{Q}|$ to ensure that any magnetic signal becomes vanishingly small and can be ignored. In principle, $\rm FR^o_{\alpha}$ should depend neither on temperature, nor on samples or on the direction of the neutron polarization. However, empirically, $\rm FR^o_{\alpha}$ would typically depend on the  studied  sample and on the direction $\alpha$ of the neutron polarization and would also display a slight monotonic variation as a function of temperature. This is basically due to inhomogeneities of the polarization within the neutron beam. Cooling and warming cause small displacements of the sample due to the thermal contraction of the stick holding the sample and attached to the cold head of the closed-circle refrigerator. Similarly, $\rm FR^o_{\alpha}$ can be sample dependent due to different sample mosaics and shapes. Both effects in an inhomogeneous polarized neutron beam induce slight changes of $\rm FR^o_{\alpha}$.
It is worth emphasizing that, in the worse case, $\rm FR^o_{\alpha}$ can  exhibit a variation of a few percents in the temperature range between $\rm T_c$ and room temperature. This T-dependence has to be considered in order to discriminate a magnetic signal from the polarization leakage, but have a secondary impact on the determination of the intrinsic T-dependence of the magnetic signal. Indeed, it implies a correction of a few percents, much smaller than the experimental error bars. 

Fig.~\ref{PRB2-Fig3}.a-d shows the temperature dependencies of the inverse flipping ratio measured at the Bragg reflections (1,0,L) for increasing odd L values. The data for the UD-85 and OD-70 samples are superimposed in the same figure. This direct comparison between both samples is allowed since the bare flipping ratios for each sample (corresponding to the polarization leakage) are basically T-independent in both cases (Fig.~\ref{PRB2-Fig3}.e) and reduce to a simple offset when comparing both samples \cite{DeAlmeida-PRB12}. For the OD sample, $\rm 1/ FR_x (T)$ remains featureless at any L values, confirming that no sizeable magnetic signal can be detected by polarized neutron scattering in that sample. On the contrary, a magnetic signal appears below $\sim$230 K at L=1 and L=3. It is worth indicating that the data reported in Fig.~\ref{PRB2-Fig3}.b and Fig.~\ref{PRB2-Fig4}.a at L=3 correspond to two distinct measurements, pointing out that the results are perfectly reproducible. Increasing further L to 5 and 7, the magnetic signal vanishes. Searches for the existence of a magnetic signal at even integer L values or for non integer L  have remained unsuccessful. This study emphasizes that the magnetic order preserves the lattice translation invariance and exhibits 3D correlations. In agreement with the data available in the literature, additional magnetization measurements on smaller UD-85 single crystals do not show any indication for a ferromagnetic (parasitic) order \cite{Leridon-Bi2212}.


\subsubsection*{\label{Polar-analysis} Polarization analysis }

\begin{figure}[t]
\includegraphics[width=6.5cm,angle=0]{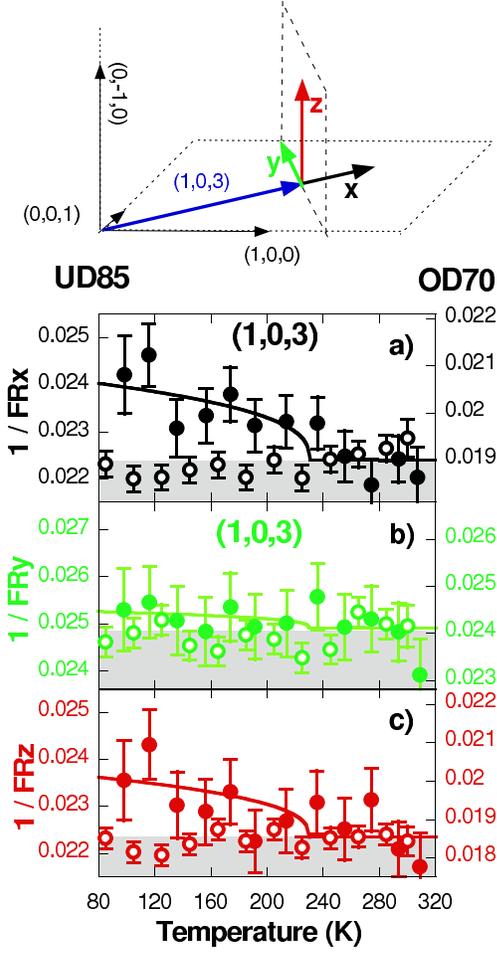}
\caption {
(color online) (Top panel) Orientation of the three orthogonal polarizations $\rm H{\alpha}$ ($\rm \alpha = (x,y,z)$) with respect to the scattering plane $\rm [1,0,0]$/$\rm [0,0,1]$ and the transferred wave vector $\rm {\bf Q}$. a-c) Temperature dependencies of the inverse flipping ratio $\rm 1/ FR_{\alpha}$ for the UD-85 sample  (left axis, full symbols) and OD-70 (right axis, opened symbols) for different polarizations: a) $\rm Hx$ (black), b) $\rm Hy$ (green), c) $\rm Hz$ (red). Solid lines are guides to the eye.
}
\label{PRB2-Fig4}
\end{figure}

In the previous sections, we exclusively studied the scattered intensity in the SF channel for the polarization Hx, which gives access to the full magnetic scattering. We now compare the scattered intensity in the SF channel for the 3 orthogonal polarizations Hx, Hy and Hz. The top panel in Fig.~\ref{PRB2-Fig4} shows the orientation of each of the 3 polarizations with respect to the scattering plane. In the absence of chirality, the polarization sum rule implies that  $\rm I_{magx}$ has to be equal to the sum of $\rm I_{magy}$ and $\rm I_{magz}$ in the SF channel. Fig.~\ref{PRB2-Fig4} reports the temperature dependence of $\rm 1/FR_{\alpha}(T)$ measured for the 3 polarizations in the OD-70 and UD-85 samples. The measurements are performed at the Bragg reflection (1,0,3). This reflection in Bi2212 is close to the reflection (1,0,1) in Hg1201 and Y123. For the OD-70 sample, $\rm 1/FR_{\alpha}(T)$ remains featureless and independent of the selected polarization $\rm H \alpha$ (Fig.~\ref{PRB2-Fig4} a-c). In contrast in the UD-85 sample, $\rm 1/FR_{x}(T)$ starts increasing below $\rm T_{mag} \sim$230 K (Fig.~\ref{PRB2-Fig4}.a). Rotating the polarization, one finds that $\rm 1/FR_{z}(T)$ displays a similar temperature dependence whereas $\rm 1/FR_{y}(T)$ is almost T-independent. This indicates that the enhancement of the scattering intensity in the SF channel depends on the neutron spin polarization, as expected for a magnetic scattering. One can further estimate that at least 3/4 of the magnetic scattering remains in the SF channel for Hz, and at most 1/4 is left for Hy. The polarization sum rule seems therefore to be fulfilled with a minimum balance factor $\rm R= I_{magz} / I_{magy} \sim$ 3. This is at variance with the polarization analysis carried out in Hg1201 and Y123 at (1,0,1) where $\rm R \sim$ 1. This may also indicate that the magnetic moments in Bi2212 are predominantly perpendicular to the ${\rm CuO_2}$ plane as it is expected in the LC model\cite{CC-review,Varma06}.

\subsubsection*{\label{co-planar} Co-planar magnetic model for bilayer systems}

More specifically and following ref.~\cite{CC-review}, one can consider a set of two staggered magnetic moments, located in  the $\rm CuO_2$ unit cell at equal distance displaced from a Cu site along the $\rm [1,1,0]$ direction. Each moment is characterized by its vertical magnetic component $\rm M_c=M cos(\theta)$ and its planar magnetic component $\rm M_{ab}=M sin(\theta)$ and $\theta$ stands for the tilt angle of the magnetic moment with respect to the $\rm {\bf c}$ axis. $\rm M$ is the average ordered magnetic moment, corresponding to the order parameter.
Further, one neglects here the possible effect of the orthorhombic distortion and considers an isotropic planar component $\rm M_{ab}$, assuming that the directions (1,0)  and (0,1) are equivalent. The magnetic signal measured at $\rm {\bf Q}$=(1,0,L) then reads:
\begin{eqnarray}
I_{magx} &=& I_{magy} + I_{magz} \\
I_{magy} &\propto& |F_m|^2 |f(L)|^2 [ \frac{1}{2} M_{ab}^2 ]   \\
I_{magz} &\propto& |F_m|^2 |f(L)|^2 [ \frac{q_l^2}{2} M_{ab}^2 + (1- q_l^2) M_c^2]
\label{Polar-dep}
\end{eqnarray}
with $\rm q_l= \frac{2 \pi}{c} L /|{\bf Q}|$. $\rm |F_m|$ stands for an effective form factor, characterizing the spatial extension of the magnetic moment. $\rm |f(L)|$ corresponds to the magnetic structure factor within the unit cell along the $\rm(\bf c)$ axis. For a monolayer system, it is equal to unity, whereas for a bilayer system it is dependent on the arrangement of the magnetic moments within the bilayer. According to Eq.~\ref{Polar-dep}, the tilt angle $\rm \theta$ is defined as 
$\rm tan(\theta)=\frac{M_{ab}}{M_c}=\sqrt{\frac{2(1-q_l^2)}{R-q_l^2}}$. For the minimum value of $\rm R$ of about 3 as deduced from Fig.~\ref{PRB2-Fig4} (R$\gtrsim$ 3), $\rm \theta$ reaches a typical value of $\rm 20\pm20^{o}$. In the other bilayer system Y123, $\rm \theta$ is found to be $\rm 35 \pm 7^{o}$ at (1,0,0) and $\rm 55 \pm 7^{o}$ at (1,0,1) in the most accurate experiment on YBa$_2$Cu$_3$O$_{6.6}$\cite{Mook}, yielding a conservative estimate of  $\rm \theta  = 45 \pm 20^{o}$  valid over few samples and Bragg spots \cite{CC-review}. The data in Hg1201 and La214 lead to the same tilt of the moment relative to the {\rm c} axis \cite{CC-review,Li-PRB,Baledent-LSCO}. The tilt angle in Bi2212 seems reduced in comparison with the average tilt angle found in Y123, Hg1201 and La214.


\subsubsection*{\label{hole-doping} Evolution of the IUC magnetic order close to optimal doping}

In Y123 \cite{Fauque} and Hg1201 \cite{Li-Nature},  the variation as a function of the hole doping (p) of the magnetic ordering temperature $\rm T_{mag}$ matches the evolution of $\rm T^{\star}$ determined from resistivity measurements. Both temperatures decrease linearly upon increasing the hole doping, at least from p$\sim$0.09 up p$\sim$0.13-0.14, corresponding to regime I in 
 Fig.~\ref{PRB2-Fig0}. The linear extrapolation of $\rm T_{mag}$ at larger hole doping suggests that the IUC magnetic order is likely to vanish around a critical hole doping $\rm p_c \sim 0.19$, considered as the end point of the PG state according to thermodynamic measurements. In Bi2212, the IUC magnetic order has been detected at rather high temperature in the normal state of the OD-87 sample \cite{DeAlmeida-PRB12}. 

\begin{figure}[t]
\includegraphics[width=6.5cm,angle=0]{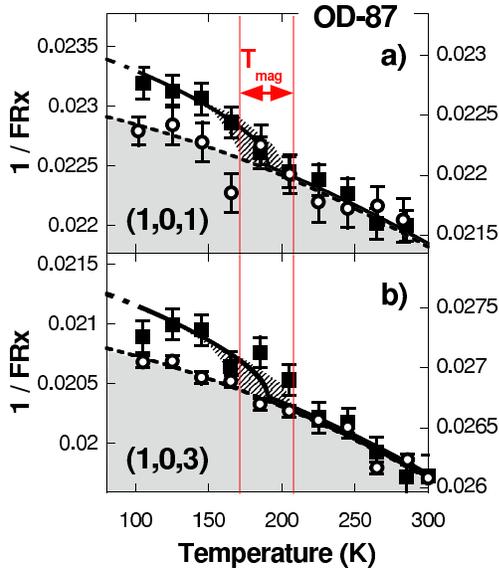}
\caption {
(color online) Bi2212 OD-87 sample  : $\rm 1/FRx(T)$ measured at the wave vectors $\rm {\bf Q}$=(1,0,L) (left axis, full symbols) and $\rm {\bf Q}$=(2,0,0) (right axis, open symbols) : a) L=1, b) L=3.
}
\label{PRB2-Fig5}
\end{figure}

This early study focussed on a unique Bragg reflection $\rm {\bf Q}$=(1,0,1), suggesting a $\rm T_{mag}$ of $\sim$170 K. We re-investigate here the IUC magnetic order in the OD-87 sample, considering the magnetic intensity at the Bragg reflections $\rm {\bf Q}$=(1,0,1) and (1,0,3). Fig.~\ref{PRB2-Fig5}.a-b  show $\rm 1/FR_{x} (T)$  at $\rm {\bf Q}$=(1,0,1) and (1,0,3) respectively. $\rm FR^o_{x}$ is given by the measurement of $\rm 1/FR_{x} (T)$  at $\rm {\bf Q}$=(2,0,0). At variance with the UD-85 and OD-70 samples, $\rm FR^o_{x}$ depends on temperature and  increases on cooling down. It is worth pointing out that the studies of the Bragg reflection (1,0,1) and (1,0,3) were carried out during two distinct experiments, but the slopes of  $\rm FR^o_{x}$  are similar in  Fig.~\ref{PRB2-Fig5}.a and Fig.~\ref{PRB2-Fig5}.b. At $\rm {\bf Q}$=(1,0,1) and (1,0,3), a magnetic signal develops around $\rm T_{mag}=190 \pm 20$K  but displays a magnitude of only $\sim $5 10$\rm ^{-4}$ weaker than the underlying nuclear Bragg intensity. When increasing the hole doping from  the UD-85 sample  to  the OD-87 sample, $\rm T_{mag}$ hardly reduces  from 230 K to 190 K. That confirms that $\rm T_{mag}$ flattens in the regime II as it was found for $\rm T^{\star}$ in ARPES data \cite{Vishik} (Fig.~\ref{PRB2-Fig0}). However, the magnetic intensity drops down by a factor $\sim$ 3 in that regime. 

\begin{figure}[t]
\includegraphics[width=7cm,angle=0]{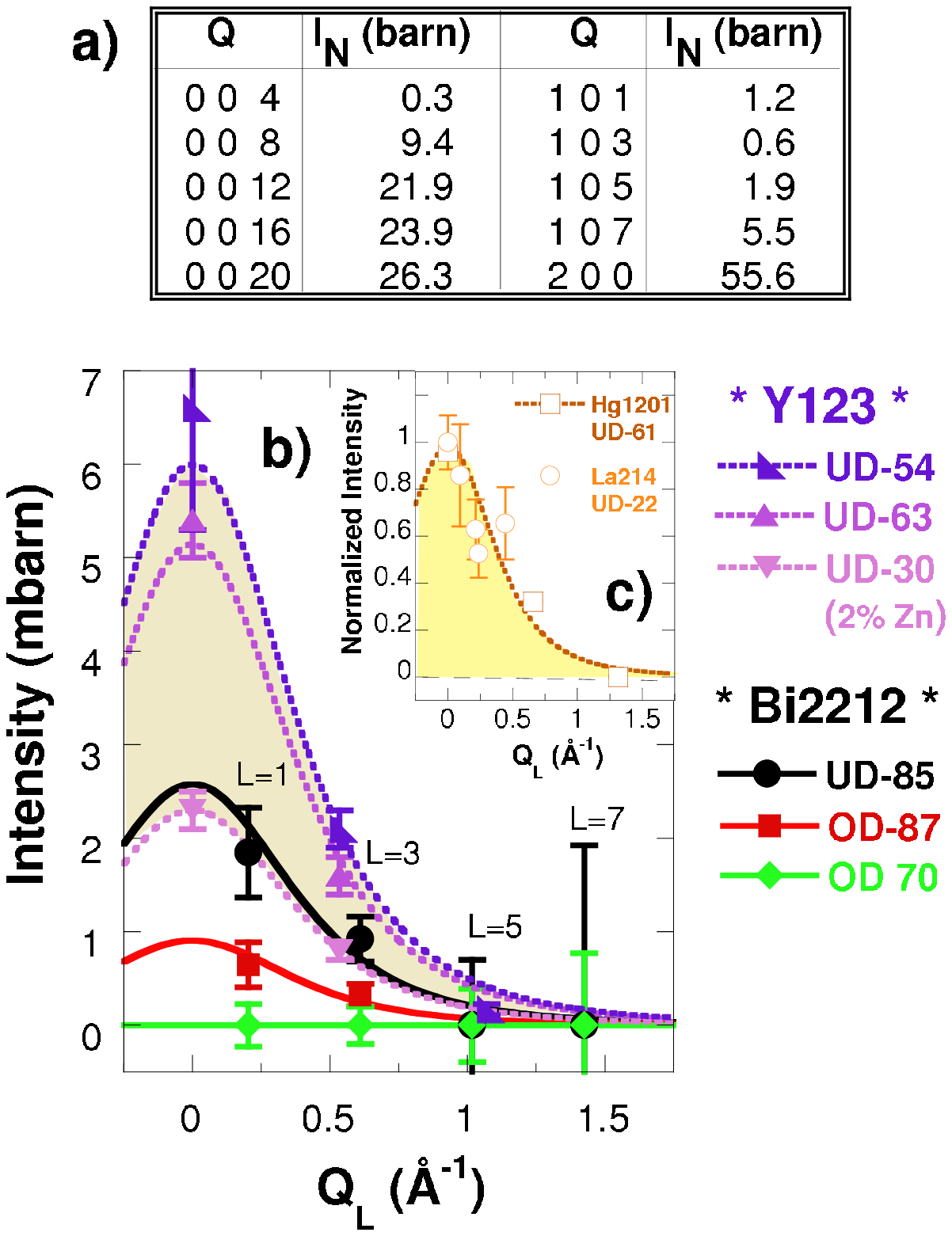}
\caption {
(color online) a) Calibration of the nuclear Bragg intensities in absolute unit (barn) for  Bi2212 samples. b) Magnetic intensity at Bragg wave vectors, measured in the normal state at $\rm T_c$+25 K, as a function of $\rm Q_L=\frac{2 \pi}{c}$ in Bi2212 (UD-85, OD-87, OD-70) and in Y123 (UD-54 \cite{Fauque,Sidis}, UD-63 \cite{Mook}, UD-30-2\%Zn \cite{Baledent-YBCO}). c) $\rm Q_L$ dependence of the normalized magnetic intensity in the monolayer cuprates Hg1201 (UD-61 \cite{Li-Nature}) and La214 (UD-22 \cite{Baledent-LSCO}). Lines correspond to a unique function for which the intensity varies only (see text).
}
\label{PRB2-Fig6}
\end{figure}
\subsubsection*{\label{Calibration} Calibration of the magnetic intensity}

In neutron diffraction, the magnetic intensity can be given in absolute unit (barn). To estimate that calibration, one just needs to determine the nuclear Bragg intensity in absolute unit. The nuclear scattering cross-section per unit cell $\rm \frac{1}{N}\frac{\partial \sigma}{\partial \Omega}_{nucl.}$ is given by\cite{squires}:
\begin{equation}
\frac{1}{N}\frac{\partial \sigma}{\partial \Omega}\large{|}_{nucl.} =  \frac{(2 \pi)^3} {v_o} \sum_{\bf G} \delta({\bf Q}-{\bf G}) |F_N|^2
\end{equation}
where $v_o$ is the volume of the unit cell and $\rm N$ the number of unit cells within the sample, i.e the sample volume $\rm V$ divided by $v_o$. $\rm {\bf G}$ is a wave vector of the reciprocal lattice. The nuclear structure factor $\rm F_N$ reads:
\begin{equation}
\label{FN}
F_N = \sum_n b_n e^{i {\bf Q}. {\bf R}_n}
\end{equation}
where $b_n$ and $\rm {\bf R}_n$ correspond to the neutron scattering length and the position of the atom n in the unit cell.

In most cuprates, that nuclear structure factor can be readily computed as all atomic positions are perfectly known.  
In contrast, in Bi2212, the strong incommensurate modulation of the $\rm BiO$ plane with respect to the $\rm CuO_2$ plane makes the determination of the nuclear structure factor extremely difficult at a quantitative level \cite{janine} as all atomic positions, $\rm {\bf R}_n$, are strongly displaced from their average position. To overcome this difficulty, one can  rather measure
$\rm \frac{1}{N}\frac{\partial \sigma}{\partial \Omega}_{nucl.}$ than compute this quantity. We then measured
the nuclear intensities of Y123 samples where the structure factors can be easily computed using Eq. \ref{FN} to calibrate the flux of the spectrometer.

To carry out this calibration, we used an $\rm YBa_2Cu_3O_7$ sample (V=3 $\rm cm^3$, mosaic=$\rm 1.3^{o}$). The sample was aligned in the [100]/[001] scattering plane on the thermal diffractometer 3T1 at reactor Orph\'ee (Saclay). The incident wave vector was set to 2.662 ~\AA$\rm ^{-1}$ and  pyrolytic graphite filters were inserted in the direct beam to remove higher order harmonics. With this experimental set-up, the intensities of a large number of Bragg reflections were collected  with a counting time of 2.5s per points. The measured Bragg intensity at a given Bragg reflection was then compared to the computed nuclear scattering cross-section convoluted with the instrumental resolution function. In order to crosscheck this calibration procedure, a second sample, $\rm YBa_2Cu_3O_{6.6}$ (V=1.7 $\rm cm^3$, mosaic=$\rm 1.3^{o}$) was also measured.

Using this procedure, one obtains a conversion factor between a number of counts in the detector per second to a number of barns times N the number of unit cells of the sample. Then, the nuclear Bragg reflections of our UD-85 sample (V$\simeq$0.05 $\rm cm^3$, mosaic=$\rm 1.3^{o}$) were measured on the same spectrometer with the same experimental set-up. Since this calibration procedure relies on an accurate knowledge of the sample volume, we used a second Bi2212 sample to crosscheck our calibration. This second sample is a small single crystal (V=0.05 $\rm cm^3$, mosaic=$\rm 0.8^{o}$) extracted from our co-aligned  Bi2212 OD-70 sample. For both samples, the intensities of the main nuclear Bragg reflections are consistent within error bars ($\rm \pm 10 \% $). The nuclear structure factors thus obtained in Bi2212 are given in barn in the table reported in Fig.\ref{PRB2-Fig6}.a for a few Bragg peaks.

The magnetic intensity is next obtained from the ratio of the magnetic contribution from the nuclear intensity at the Bragg peaks (1,0,L). For L=1 and 3, this is given by the enhancement of $\rm 1/ FR_x (T)$ upon approaching $\rm T_c$: it is of the order of $\rm \sim$ 1.5 $\rm 10^{-3}$ for the UD-85 sample. The magnetic intensities, as reported in Fig.~\ref{PRB2-Fig3}.(a-b), are then given in absolute unit as $\rm I_{mag} \sim$1.8 mbarn at $\rm {\bf Q}$=(1,0,1) and $\sim$0.9 mbarn at $\rm {\bf Q}$=(1,0,3). Following the same calibration in absolute units, one can estimate the magnetic intensities  for the OD-87 sample: 0.6$\pm$0.3 mbarn  at $\rm {\bf Q}$=(1,0,1) and 0.3$\pm$0.1 mbarn at $\rm {\bf Q}$=(1,0,3). Fig.\ref{PRB2-Fig6}.b shows the variation along the $\rm [001]$ direction of the magnetic intensity in the normal state just above $\rm T_c$ (typically measured at $\rm T_c +$25 K), in Bi2212 and Y123 samples (UD-54: $\rm T_{mag}=300 \pm 10$ K \cite{Fauque,Sidis}, UD-63: $\rm T_{mag}=235 \pm 15$ K \cite{Mook}, UD-30-2\%Zn: $\rm T_{mag}=250 \pm 20$ K \cite{Fauque,Sidis}). This variation can be described by a unique phenomenological form (here a squared Lorentzian function), where only the amplitude varies from one sample to another. While all bilayer samples display the same fast decay along the $\rm [001]$ direction, the measured intensity vary significantly for samples with comparable $\rm T_{mag}$.

In the bilayer system Y123 \cite{Fauque,Sidis}, the fast decay of the magnetic Bragg intensity along c* was first ascribed to a ferromagnetic coupling of the magnetic moments within the bilayer as the L=0 peak displays the largest magnetic contribution. In such a model, the magnetic structure factor is indeed weighted by a term in Eq. \ref{Polar-dep} like $f(L)=2 \cos(\pi\frac{d}{c}L)$, where $\rm d$=3.3~\AA$ $  corresponds to the distance between $\rm CuO_2$ planes within the bilayer. However, the magnetic intensity in the monolayer systems Hg1201 \cite{Li-Nature} and La214 \cite{Baledent-LSCO} is found surprisingly to display the same fast decay (Fig.\ref{PRB2-Fig6}.c), although the bilayer structure is absent. The scaling of the magnetic intensity measured at $\rm {\bf Q}$=(1,0,L) in four distinct cuprates families \cite{DeAlmeida-PRB12} actually suggests that this decay could be a generic feature of the observed magnetic order which actually would not depend on the coupling within the $\rm CuO_2$ bilayer.

As we did  in Y123 \cite{Fauque,Sidis,Mook,CC-review,CC-review2}, we can next give a rough estimate of the ordered moment. In ref.~\cite{CC-review}, we have shown that in Y123, 1 mbarn at (1,0,1) corresponds to an ordered moment of $\rm M \sim 0.1 \mu_B$ under some simple assumptions on the magnetic form factor. Similarly, in Bi2212, 1 mbarn at (1,0,3) would also correspond to $\rm M \sim 0.1 \mu_B$. In the insulating antiferromagnetic (AFM) cuprates, the staggered magnetic moment is typically ~5 times larger \cite{Rossat}. Since the magnetic intensity is proportional to the square of the ordered moment, the magnetic intensity is then typically  1 to 2 orders of magnitude weaker than in an insulating AFM parent compound around (0.5,0.5,L). However, it is worth pointing out that the AFM occurs at a different wave vector. Further, it corresponds to the ordering of Cu S=1/2 spins only whereas the IUC (Q=0) magnetic order, that we are reporting here, cannot be described by Cu spins only. 

\section{\label{discussion} Discussion and concluding remarks}

\begin{figure}[t]
\includegraphics[width=8cm,angle=0]{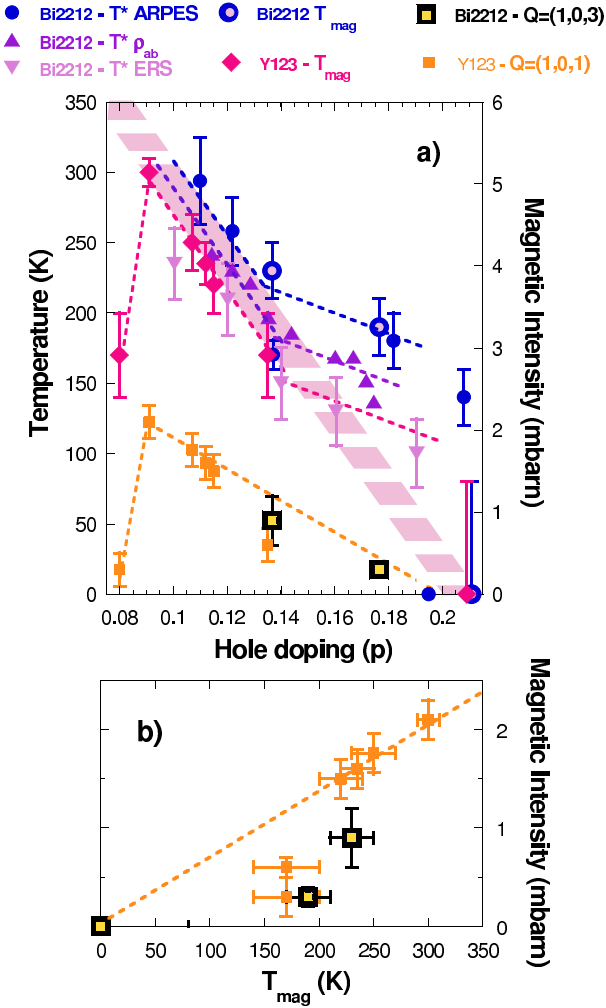}
\caption {
(color online) a) Left axis: hole doping dependence of the pseudo-gap (PG) temperature $\rm T^{\star}$ in Bi2212, measured by ARPES \cite{Vishik}, ab-resistivity \cite{Raffy} and ERS \cite{Sacuto}. Hole doping dependencies of the characteristic temperature associated with the appearance of the IUC magnetic phase $\rm T_{mag}$  for the bilayer systems Bi2212 and for Y123 \cite{Fauque,Mook,Baledent-YBCO}. Right axis: variation of magnetic Bragg intensity measured at $\rm {\bf Q}$= (1,0,L) with: L=3 for Bi2212 ($\rm |Q|$=1.75 ~\AA$\rm^{-1}$), L=1 for Y123 ($\rm |Q|$=1.71 ~\AA$\rm^{-1}$). b) Evolution of the magnetic intensity as a function of $\rm T_{mag}$.
}
\label{PRB2-Fig7}
\end{figure}

Fig.~\ref{PRB2-Fig7}.a summarizes the variation of the PG temperature $\rm T^{\star}$ in Bi2212. The hole doping level, p,  is given by the  phenomenological relationship \cite{Loram}: $\rm T_c=T_{c,max} [1-82.6(p-0.16)^2]$. $\rm T^{\star}$ is determined by three different techniques: ARPES \cite{Vishik} ($\rm T_{c,max}$=96 K), ab-resistivity \cite{Raffy} ($\rm T_{c,max}$=82 K), ERS \cite{Sacuto} ($\rm T_{c,max}$=90 K). Within a range of $\rm \delta p \pm $0.01 and $\rm \delta T \pm $25 K, the $\rm  T^{\star} (p)$ values given by these three techniques are consistent within error bars. In the T-p phase diagram, one can observe a $\rm  T^{\star} (p)$ band, rather than a single line of transition. The distribution of $\rm  T^{\star}$ values within this band reflects the fact that each technique develops its own criteria to estimate $\rm  T^{\star}$.

 Fig.~\ref{PRB2-Fig7} shows that the  variation of $\rm  T_{mag}$ as a function of hole doping exhibits the same trends for the bilayer systems Bi2212 and Y123. For Bi2212, $\rm  T_{mag} (p)$ matches quite well $\rm  T^{\star} (p)$  determined by ARPES \cite{Vishik} but $\rm  T^{\star} (p)$  determined by ERS \cite{Sacuto} occurs at a slightly lower temperature. In the hole doping range between p$\sim$0.09 and p$\sim$0.13-0.14 (regime I), the PG temperature decreases linearly along a single $\rm  T^{\star}(p)$ band (Fig.~\ref{PRB2-Fig7}.a). At larger hole doping (regime II), the $\rm  T^{\star}$ decay slows down and the discrepancies between the $\rm  T^{\star}$ values provided by different techniques increase. $\rm  T_{mag} (p)$ in Y123 seems to correspond to the lower bound of the $\rm  T^{\star}(p)$ band. Since Y123 system is usually considered as a much cleaner system that Bi2212 system, one can speculate that interstitial oxygen dopants and vacancies at the apical oxygen in Bi2212 \cite{Hoffman} induce some disorder within the $\rm CuO_2$ planes which broadens the PG transition in a wide hole doping range and allows the persistence of spatially reduced PG phases even at large doping.

When dealing with a long range magnetic order in a homogeneous system, the intensity at a magnetic Bragg reflection is proportional to $\rm M^2$, the square of the ordered moment. Within a mean field approach, the ordering temperature is also expected to scale with $\rm M^2$.  As a consequence, one may expect the magnetic intensity to scale with $\rm T_{mag}$ when varying the hole doping in the case of a uniform long range magnetic order. In addition to $\rm T_{mag}(p)$, Fig.~\ref{PRB2-Fig7}.a shows the hole doping dependencies of the magnetic intensity measured in the normal state ($\sim$ $\rm T_c+$25 K) at wave vectors  $\rm {\bf Q}$=(1,0,L) with: L=3 for the bilayer system Bi2212 ($\rm |Q|$=1.75 ~\AA$\rm^{-1}$, $\rm Q_L$=0.61 ~\AA$\rm^{-1}$), L=1 for the bilayer system Y123 \cite{Fauque,Mook,Baledent-YBCO} ($\rm |Q|$=1.71~\AA$\rm^{-1}$, $\rm Q_L$=0.54 ~\AA$\rm^{-1}$). Since the $\rm Q_L$ values for both bilayer systems are slightly different, the magnetic intensities in Y123 could be overestimated by $\rm \sim 25$\% with respect to those in Bi2212, owing to the fast decay of the magnetic intensity along {\bf c}$^{\star}$  (Fig.~\ref{PRB2-Fig6}.b). As shown Fig.~\ref{PRB2-Fig7}.a-b, the magnetic intensities  scale with the $\rm T_{mag}$ values in a hole doping range corresponding to regime I. Outside this hole doping range, the magnetic intensity interestingly decreases much faster than $\rm T_{mag}$ (regime II).

Why this can be the case ? A scaling relationship between the magnetic intensity and the ordering temperature can typically break down when the magnetic correlation length shortens, yielding a redistribution of the magnetic intensity in momentum space around the magnetic Bragg wave vector. For instance, this has been discussed in the context of Cu spins AFM ordering at very low doping in cuprates. Indeed, in lightly doped  Y123 \cite{Rossat} and La214 \cite{AF-La214}, the N\'eel temperature does not scale with $\rm M^2$ upon doping. The scaling relationship also breaks down when the system becomes inhomogeneous and the magnetic order occupies only a reduced fraction of the sample. In the case of the IUC magnetic order, the limitation of $\rm \xi_{ab}$ and $\rm \xi_c$, the in-plane and out-of-plane magnetic correlation lengths, can account for the breakdown of the scaling relationship between Bragg magnetic intensity and $\rm T_{mag}$.

In Y123 and Bi2212,  the observation of the magnetic Bragg intensity highlights only the existence of 3D magnetic correlation lengths and, so far, there is no direct evidence of finite correlation lengths by diffraction. A limited reduction of the magnetic correlation length is actually difficult to observe directly. Indeed, the neutron resolution is relaxed in order to maximize the scattered neutron intensity. For a resolution limited Bragg scattering, one can at best estimate a lower bound of $\sim $75 ~\AA ~for $\rm \xi_{ab}$ and $\rm \xi_c$. In contrast, a strong reduction of the magnetic correlation length can be observed directly. For instance, this has been demonstrated in lightly doped La214 \cite{Baledent-LSCO}, where the IUC magnetic order is 2D and $\rm \xi_{ab} \sim$2-3 a. 

In Bi2212,  the fast decay of the magnetic scattering intensity  from the  UD-85 sample to the OD-87 sample can then be understood by a weakening of the magnetic correlation length. The OD-87 sample  is close to critical hole doping at which the PG state is expected to vanish. As the PG state is a broken symmetry state ending at a quantum critical point, the OD-87 sample  lies within a quantum critical regime, largely controlled by fluctuations. Keeping in mind that Bi2212 is a rather disordered material, one can speculate that a certain quenched disorder could pin down and freeze the critical fluctuations at high temperature around the quantum critical point. Within this scenario, the intrinsic disorder of a given cuprate family, could allow the finite size PG domains to survive at rather high temperature around optimal doping. This argument applies particularly well for the orbital LC model having an Ising-like discrete symmetry \cite{Varma06,RFIM}. The existence of a quenched disorder would block the quantum critical fluctuations associated with the LC order.

Another interesting related situation is given by Zn substitution in underdoped Y123. While Zn substitution is known to preserve the hole doping level, $\rm T_{mag}$ is found unchanged but the magnetic Bragg intensity drops down by a factor $\rm \sim 2$ with respect to a Zn-free sample \cite{Mook} in a $\rm YBa_2(Cu_{1-y}Zn_{y})_3O_{6.6}$ sample \cite{Baledent-YBCO} (UD-30, y=2\%) (see Fig.~\ref{PRB2-Fig6}). Zn induces a disorder which likely reduces the volume fraction of the sample occupied by the IUC magnetic order. 
By analogy, the presence of larger disorder in Bi2212 may explain why the magnetic intensity is found weaker in Bi2212 than Y123 for a given $\rm T_{mag}$ (Fig.~\ref{PRB2-Fig7}.b).

Our previous polarized neutron scattering  study  of Bi2212 system \cite{DeAlmeida-PRB12}  provided evidence for the existence of an IUC magnetic order in the PG state of this system. The magnetic order could be observed  on the Bragg reflection  $\rm {\bf Q}$=(1,0,1) for two samples, UD85 and OD87. In the present study, we confirm this observation on the Bragg reflection  $\rm {\bf Q}$=(1,0,3). The full polarization analysis performed on the UD-85 sample  allows an estimate of the tilt angle of the magnetic moments with respect to the c-axis ($\rm \theta=20\pm20^{o}$). This angle is significantly smaller than the value $\rm \theta= 45\pm20^{o}$ found for the bilayer system Y123 and the monolayer systems Hg1201 and La214. In Bi2212, the evolution of $\rm T_{mag}$ as a function of the hole doping matches the evolution of $\rm T^{\star}$ reported by various techniques \cite{DeAlmeida-PRB12} and, in particular ARPES measurements \cite{Vishik}. The persistence of $\rm T_{mag} \simeq T^{\star}$, as high as $\sim$190 K,  slightly above optimal doping is a striking feature. While $\rm T_{mag}$ weakly decreases from p$\sim$0.13-0.14 to p$\sim$0.18, the magnetic intensity drops down by a factor 3. After a calibration in absolute unit using a two-step calibration procedure (different from the self-calibration procedure  used for other cuprates families \cite{CC-review}), the magnetic intensities for Bi2212 have been compared with those reported for Y123. This comparison reveals that the ordering temperature does not scale with the square of the ordered magnetic moment, as expected in a mean field theory. This effect suggests that the range of magnetic correlations might be finite at large doping.

\paragraph*{Acknowledgments.} We wish to thank  Mun Chan, Seamus Davies, Yann Gallais, Martin Greven, Brigitte Leridon and Chandra Varma for stimulating discussions on various aspects related to this work.


\end{document}